\documentstyle[epsf]{mn}

\newif\ifAMStwofonts

\ifoldfss
  \ifCUPmtlplainloaded \else
    \NewTextAlphabet{textbfit} {cmbxti10} {}
    \NewTextAlphabet{textbfss} {cmssbx10} {}
    \NewMathAlphabet{mathbfit} {cmbxti10} {} 
    \NewMathAlphabet{mathbfss} {cmssbx10} {} 
  \fi
  \ifAMStwofonts
    \ifCUPmtlplainloaded \else
      \NewSymbolFont{upmath} {eurm10}
      \NewSymbolFont{AMSa} {msam10}
      \NewMathSymbol{\upi}     {0}{upmath}{19}
      \NewMathSymbol{\umu}     {0}{upmath}{16}
      \NewMathSymbol{\upartial}{0}{upmath}{40}
      \NewMathSymbol{\leqslant}{3}{AMSa}{36}
      \NewMathSymbol{\geqslant}{3}{AMSa}{3E}

       \let\le=\leqslant
       \let\ge=\geqslant
    \fi
  \fi
\fi 

\ifnfssone
  \newmathalphabet{\mathit}
  \addtoversion{normal}{\mathit}{cmr}{m}{it}
  \addtoversion{bold}{\mathit}{cmr}{bx}{it}
  \newmathalphabet{\mathbfit} 
  \addtoversion{normal}{\mathbfit}{cmr}{bx}{it}
  \addtoversion{bold}{\mathbfit}{cmr}{bx}{it}
  \newmathalphabet{\mathbfss} 
  \addtoversion{normal}{\mathbfss}{cmss}{bx}{n}
  \addtoversion{bold}{\mathbfss}{cmss}{bx}{n}
  \ifAMStwofonts
    \ifCUPmtlplainloaded \else
      \UseAMStwoboldmath
      \makeatletter
      \new@mathgroup\upmath@group
      \define@mathgroup\mv@normal\upmath@group{eur}{m}{n}
      \define@mathgroup\mv@bold\upmath@group{eur}{b}{n}
      \edef\UPM{\hexnumber\upmath@group}
      \new@mathgroup\amsa@group
      \define@mathgroup\mv@normal\amsa@group{msa}{m}{n}
      \define@mathgroup\mv@bold\amsa@group{msa}{m}{n}
      \edef\AMSa{\hexnumber\amsa@group}
      \makeatother
      \mathchardef\upi="0\UPM19
      \mathchardef\umu="0\UPM16
      \mathchardef\upartial="0\UPM40
      \mathchardef\leqslant="3\AMSa36
      \mathchardef\geqslant="3\AMSa3E

       \let\le=\leqslant
       \let\ge=\geqslant
    \fi
  \fi
\fi 

\ifnfsstwo
  \DeclareMathAlphabet{\mathbfit}{OT1}{cmr}{bx}{it}
  \SetMathAlphabet\mathbfit{bold}{OT1}{cmr}{bx}{it}
  \DeclareMathAlphabet{\mathbfss}{OT1}{cmss}{bx}{n}
  \SetMathAlphabet\mathbfss{bold}{OT1}{cmss}{bx}{n}
  \ifAMStwofonts
    \ifCUPmtlplainloaded \else
      \DeclareSymbolFont{UPM}{U}{eur}{m}{n}
      \SetSymbolFont{UPM}{bold}{U}{eur}{b}{n}
      \DeclareSymbolFont{AMSa}{U}{msa}{m}{n}
      \DeclareMathSymbol{\upi}{0}{UPM}{"19}
      \DeclareMathSymbol{\umu}{0}{UPM}{"16}
      \DeclareMathSymbol{\upartial}{0}{UPM}{"40}
      \DeclareMathSymbol{\leqslant}{3}{AMSa}{"36}
      \DeclareMathSymbol{\geqslant}{3}{AMSa}{"3E}

       \let\le=\leqslant
       \let\ge=\geqslant
    \fi
  \fi
\fi 

\ifCUPmtlplainloaded \else
  \ifAMStwofonts \else 
    \def\upi{\pi}
    \def\umu{\mu}
    \def\upartial{\partial}
  \fi
\fi

\title[Cygnus X-2]{The Optical Light Curves of Cygnus X-2 (V1341 Cyg)
and the Mass of its Neutron Star}
\author[Orosz \& Kuulkers]{Jerome A. Orosz$^{1}$ \& 
Erik Kuulkers$^{2}$\thanks{Present address: Space Research Organization 
Netherlands, Sorbonnelaan 2, 3584 CA Utrecht, The Netherlands}\\
        $^1$Department of Astronomy \& Astrophysics, The Pennsylvania State
University, 525 Davey Laboratory, University Park, PA 16802, USA\\
$^2$Astrophysics, University of Oxford, Nuclear and Astrophysics Laboratory,
Keble Road, Oxford OX1 3RH, UK}
\date{\today}
\pagerange{\pageref{firstpage}--\pageref{lastpage}}
\pubyear{1998}

\begin{document}

\maketitle

\label{firstpage}

\begin{abstract}
We present $U$, $B$ and $V$ light curves (taken from the literature)
of the low mass X-ray binary Cygnus X-2.  We show that the most
significant photometric periods seen in the $B$ and $V$ light curves
are consistent with half of the orbital period found from spectroscopy
($P=9.8444$ days).  The ``lower envelope'' of the light curves folded
on the orbital period are ellipsoidal (i.e.\ they have two maxima and
two minima per orbital cycle).  We fit an ellipsoidal model to the
lower envelopes of the $B$ and $V$ light curves to derive inclination
constraints.  This model includes light from an accretion disc and
accounts for eclipses and X-ray heating.  Using the extreme assumption
that there is no disc light, we derive a lower limit on the
inclination of $i\ge 49^{\circ}$.  If we assume the accretion disc is
steady-state where its radial temperature profile goes as $T(r)\propto
r^{-3/4}$, we find an inclination  of $i=62.5^{\circ}\pm 4^{\circ}$.
However, the predicted ratio of
the disc flux to the total flux in $B$ (the ``disc fraction'') is
larger than what is observed ($\approx 0.55$ compared to $\le 0.3$).  If we use
a flatter radial temperature profile of the disc expected for strongly
irradiated discs ($T(r)\propto r^{-3/7}$), then we find an inclination
of $i= 54.6^{\circ}$ and a disc fraction in $B$ of
$\approx 0.30$.  However, in this
case the value of $\chi^2$ is much larger (48.4 with 36 degrees
of freedom compared to 40.9 
for
the steady-state case).  Adopting $i=62.5\pm 4^{\circ}$ and using a
previous determination of the mass ratio ($q=M_c/M_x=0.34\pm 0.04$)
and the optical mass function ($f(M)=0.69\pm 0.03\,M_{\sun}$), we find
that the mass of the neutron star is $M_x=1.78\pm 0.23\,M_{\odot}$
and the mass of the secondary star is $M_c=0.60\pm 0.13\,M_{\odot}$.
We derive a distance of $d=7.2\pm 1.1$ kpc, which is significantly
smaller than a recent distance measurement of $d=11.6\pm 0.3$ kpc
derived from an observation of a type I radius-expansion
X-ray burst, but consistent with earlier distance estimates.
\end{abstract}


\section{Introduction}
Reliable measurements of neutron star masses are important for placing
constraints on the equation of state of dense nuclear matter.  Currently,
the most precise mass measurements for neutron stars come from 
studies of binary radio pulsars (Stairs et al.\ 1998;
Thorsett \& Chakrabarty 1998). 
The masses of the neutron stars in the binary radio pulsars 
likely reflect the mass at their formation since
these particular neutron stars presumably
have not accreted any mass since their formation.
On the other hand, neutron stars in X-ray binaries have been accreting
at large rates for extended periods of time.  Hence
the mass estimates for neutron stars in X-ray binaries should give us
information on the {\em range} of allowed masses (van Kerkwijk,
van Paradijs, \& Zuiderwijk 1995a; van Paradijs 1998).   
Unfortunately,
X-ray binaries are not as ``clean'' as binary radio pulsars and mass
determinations derived from dynamical studies are subject to larger
uncertainties (e.g.\ van Kerkwijk et al.\ 1995b;   
Stickland, Lloyd, \& Radziun-Woodham 1997). 
In the case of the high-mass X-ray binaries, 
the observed radial velocity curves often show pronounced deviations
from the expected Keplerian shapes, presumably due to tidal effects
and non-radial oscillations in the high-mass secondary star.  
Since
most of the mass of the
binary resides in the high-mass O/B secondary star,
the derived neutron star
masses are quite sensitive to errors in the velocity curves
of the visible stars.
In the case of many low-mass X-ray binaries, it is often not possible
to directly observe the secondary star optically since the accretion
disc dominates the observed flux.  Hence, reliable dynamical mass estimates
are not available for many of these systems.

\begin{figure*}
\vspace{8.0cm}
\includegraphics{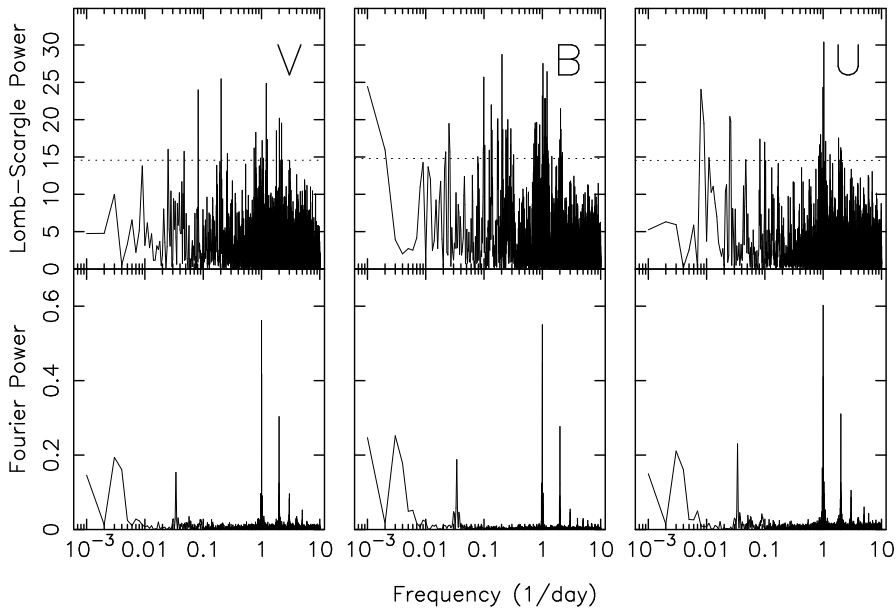}
\caption{Top panels:
The Lomb-Scargle periodograms for the entire $V$ data set
(left); the entire $B$ data set (middle), and the entire $U$
data set (right).
Bottom panels:  The Fourier transforms of the window functions of the
three data sets.}
\label{scargfig}
\end{figure*}

Cygnus X-2,
which is  one of the brightest low-mass X-ray binaries
known, is one of the rare
cases among the persistent low-mass X-ray binaries where the secondary
star is easily observed.
Cyg X-2 is known to contain a neutron star
because Type I X-ray bursts have been observed 
(Kahn \& Grindlay 1984;
Kuulkers, van der Klis, \& van Paradijs 
1995; Wijnands et al.\ 1997; Smale 1998).
The neutron star is believed to be accreting mass from its
companion at a near Eddington rate (see Smale 1998).
V1341 Cygni, its optical counterpart (Giacconi et al.\ 1967), is 
relatively bright, so reasonably precise 
spectroscopic and photometric observations can be obtained. 
The orbital period of
$P=9.844$ days was determined by Cowley et al.\ (1979) and by 
Crampton \& Cowley (1980)  
from the observed radial
velocity variations of the companion star.  
Cowley et al.\ (1979) also reported a spectral type 
for the companion star in the range of A5 to F2
(they attributed the change in the observed spectral type to X-ray
heating of the secondary). 
Casares, Charles, \& Kuulkers (1998, hereafter CCK98) have recently refined
the measurements of the orbital parameters.  They determined a period of
$P=9.8444\pm 0.0003$ days, an optical mass function of
\begin{equation}
f(M)\equiv {PK_c^3\over 2\pi G}={M_x\sin^3i\over
(1+q)^2}=0.69\pm 0.03\,M_{\sun}, 
\end{equation}
where $M_x$ is the mass of the neutron star, $M_c$ is the mass of
the companion star, $K_c$ is the semi-amplitude
of the companion's radial velocity curve, and where
$q=M_c/M_x=0.34\pm 0.04$.  CCK98 also determined a spectral type for
the companion star of A9III and reported no variation of the
spectral type with orbital phase, in contradiction to Cowley et al.\
(1979).

One needs a measurement of the orbital inclination in order to derive
mass measurements from the orbital elements of CCK98.  Models of the
optical/IR light curves are the most ``direct'' method to determine
the inclination (e.g.\ Avni \& Bahcall 1975; Avni 1978).
We have gathered 
$U$, $B$, and $V$ photometric data of Cyg X-2
from the literature with the goal of obtaining the mean light curve
and deriving the inclination.
We demonstrate that the derived 
mean orbital
light curves show the familiar signature of ellipsoidal variations.
The light curves are then modeled to place limits on the
inclination.  We also
show that the photometric period is consistent
with the spectroscopic orbital period.  We describe below the analysis
of the tabulated photometric data, period determination, and
the ellipsoidal modelling.  We conclude with a discussion of the mass
of the neutron star and the distance to the source.

\section{Cyg X-2 Light curves}

\subsection{Observations}

\begin{figure*}
\vspace{4.3cm}
\includegraphics{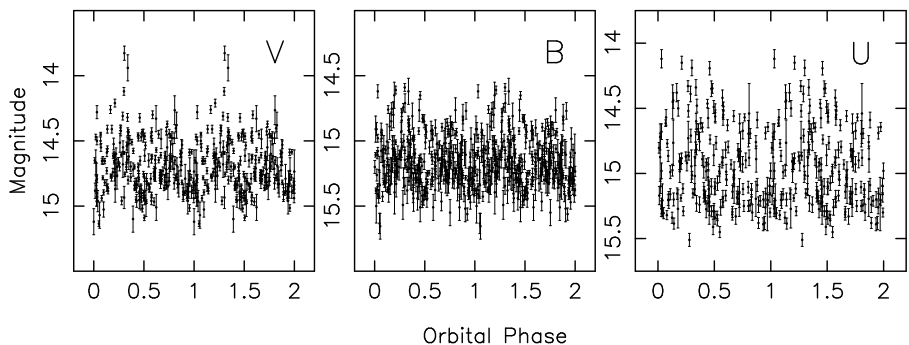}
\caption{The folded light curves for $V$ 
(left); $B$ (middle), and $U$
(right).  Each point has been plotted twice.}
\label{foldall}
\end{figure*}

We used all the photoelectric and photographic data as tabulated in the
literature for our analysis. These are photoelectric ($U$, $B$, and $V$) data 
obtained between 1967 and 1984 (Kristian et al.\ 1967; Peimbert et al.\
1968; Mumford 1970; Chevalier, Bonazzola \&\ Ilovaisky 1976; 
Lyutyi \&\ Sunyaev 1976; Ilovaisky et al.\ 1978; Kilyachkov 1978; 
Beskin et al.\ 1979; Goranskii \&\ Lyutyi 1988), 
and photographic ($B$ and $V$) data obtained in 1974 and 1975 
(Basko et al.\ 1976).
The uncertainties of the photoelectric data are typically between 
0.02--0.03 magnitudes in the $V$ and $B$ band, and 0.05--0.08 
magnitudes in the $U$ band, while the photographic data have typical 
uncertainties between 0.08--0.15 magnitudes (see Goranskii \&\ Lyutyi 
1988). Close $B$ and $V$ photographic 
magnitudes in time with $B$ and $V$ photoelectric 
magnitudes show their measurements to be consistent with each other; 
we therefore combined them.
The mean $U$, $B$ and $V$ band
magnitudes (with the rms given in brackets) of Cyg X-2 from our 
sample are 14.95 (0.31), 15.16 (0.20), and 14.70 (0.21), respectively.

\subsection{Period analysis}

Although several periodicities had been reported between $\sim$0.25 and 
$\sim$14~days before 1979, none of them were consistent with the 
orbital period as determined from the spectroscopic observations 
by Cowley et al.\ (1979) and Crampton \&\ Cowley (1980)
(see also CCK98).
After 1979 it was shown that folding the photoelectric and photographic
data on the spectroscopic period gave an ellipsoidal 
(i.e.\ double-peaked) shaped light curve 
(Cowley et al.\ 1979; Goranskii \&\ Lyutyi 1988). 
Still up to today, no period analysis of the Cyg\,X-2 light curves has given 
independent proof of the orbital variations.

We therefore subjected all the combined $U$ (469 points), $B$ 
(966 points), and 
$V$ (572 points) band data separately to 
a period analysis using various techniques (e.g., Lomb-Scargle
and phase dispersion minimization). We searched the data for periodicities
between 0.1 and 1000~days. Plots of the Lomb-Scargle periodograms can 
be found in the upper panels of Figure \ref{scargfig}. 
In the figure we also give the 3$\sigma$ confidence level, above which we regard
signals as significant. These confidence levels were determined from a
cumulative probability distribution appropriate for our three data sets 
(see e.g.\ Homer et al.\ 1996).
The most significant peak found 
in both the $B$ and $V$ band data were at a period of $\sim$4.92~days,
whereas no significant peak near this period was found in the U band data.
In the lower panels of Figure \ref{scargfig} we give the power spectra of the
corresponding window functions in the three passbands. Clearly, the most
significant peak in the U band is due to the observing window.

We estimated the error on the periods found by employing a Monte-Carlo 
technique; we generated $\sim$10,000 data sets with the same variance, 
amplitude and period as the observed data. We then subjected the faked data 
sets to the Lomb-Scargle algorithm; the distribution of the most
significant peaks then leads to 1$\sigma$ error estimates.
We also fitted a sine wave to the data 
near the found periods, where the errors in the magnitude measurements were
scaled so that the fit had a reduced $\chi^2$ of
$\sim$1; the 1$\sigma$ uncertainty in the period was determined using 
$\Delta\chi^2=1$. This resulted in similar error estimates.
We derived $P_{V} = 4.9200 \pm 0.0008$~days, and
$P_{B} = 4.9213 \pm 0.0009$~days. 
Clearly the period found in the $B$ and $V$ band data is half the 
(spectroscopic) orbital period. Our analysis therefore gives the 
first independent proof of (half) the orbital period.

No clear peak can be found near the 
$\sim 78$ day X-ray period (Wijnands, Kuulkers \&\ Smale 1996; see also
Kong, Charles \&\ Kuulkers 1998). 
However, other significant peaks are found at e.g.\ $\sim 12.1$ and 
$\sim 35.3$ days in $V$, $\sim 10.1$ and $\sim 39$ days in $B$, 
$\sim 10.1$, $\sim 12.1$, $\sim 35.3$ and $\sim 125$ days in $U$, and at 
aliases between the different periods (e.g.\ the peak at $\sim 0.82$ days
in $V$ is the alias of half the orbital period and the 
$\sim 35.3$ days period).
We note that the $\sim 35$ day period is close to the second significant 
peak in X-rays reported by Wijnands et al.\ (1996).

\subsection{Mean light curve}

Since only the $B$ and $V$ band data shows significant orbital variations,
we will concentrate only on mean light curves from these two bands.
As noted by Goranskii \&\ Lyutyi (1988), Cyg\,X-2 shows a strong concentration
of data points towards the lowest magnitudes, which they called the ``quiet
state'' (see Figure 1 of Goranskii \&\ Lyutyi 1988 and Figure \ref{foldall}). 
On top of that Cyg X-2 displays increases in 
brightness on time-scales ranging from $\approx 5$ days to
$\approx 10$ days,
flares lasting less than a day, and drops in 
brightness for a few days (Goranskii \&\ Lyutyi 1988).

As can be seen in 
Figure
\ref{foldall}, the lower envelope of the data points shows most clearly the 
ellipsoidal modulations. Rather than constructing mean light curves from data
selected from prolonged quiet states in the $B$ 
band (Goranskii \&\ Lyutyi 1988),
we used a more unbiased method.

\begin{figure}
\vspace{4.3cm}
\includegraphics{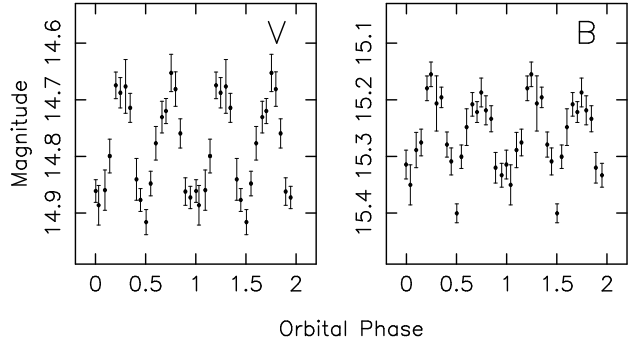}
\caption{The final folded and binned ``quiescent'' light curves for $V$
(left); and $B$ (right).  Each point has been plotted twice.}
\label{foldBV}
\end{figure}

We computed nightly averages of the observed times and magnitudes, in order
to avoid larger weights to nights with many measurements. The nightly
averages were phase 
folded on the ephemeris given by Crampton \&\ Cowley (1980),
where we used the definition for phase zero as inferior conjunction of the
companion star, which is the time at which the companion star is closest to 
us. The nightly averages were binned into 20 phase bins, and we determined the 
lower envelope of the curve by taking the first 6 lowest magnitudes per bin.
We then fitted a sine wave to the lower envelope, and subtracted it from 
the nightly averages. Finally, we discarded those data points which were
greater than 
$\approx 13$ times and $\approx 9$ times 
the rms in the mean of the whole
sine subtracted data sample,
for the $B$ and $V$ band, respectively
(there was very little change in the mean light curve
when slightly different thresholds were used).
The resulting mean $B$ and $V$ band 
folded light curves of the accepted nightly averages, 
which corresponds to the so-called ``quiescent state'', are shown in 
Figure \ref{foldBV}.  We present the folded $B$ and $V$ light curves
in tabular form in Tables 1 and 2, respectively.

\begin{table}
\centering
\begin{center}
\caption[]{The adopted folded $B$ light curve}
\begin{tabular}{ccc} \hline
Phase &    $B$ magnitude  &  magnitude error \\
\hline
0.040  &  15.350    &    0.035 \\ 
0.097  &  15.289    &    0.031 \\ 
0.147  &  15.275    &    0.024 \\ 
0.205  &  15.180    &    0.022 \\ 
0.245  &  15.155    &    0.022 \\ 
0.300  &  15.206    &    0.049 \\ 
0.349  &  15.196    &    0.018 \\ 
0.404  &  15.279    &    0.022 \\ 
0.447  &  15.309    &    0.024 \\ 
0.503  &  15.400    &    0.017 \\ 
0.548  &  15.300    &    0.021 \\ 
0.600  &  15.248    &    0.032 \\ 
0.657  &  15.208    &    0.021 \\ 
0.704  &  15.222    &    0.018 \\ 
0.746  &  15.187    &    0.025 \\ 
0.793  &  15.219    &    0.026 \\ 
0.845  &  15.234    &    0.023 \\ 
0.889  &  15.320    &    0.027 \\ 
0.950  &  15.333    &    0.021 \\ 
0.997  &  15.314    &    0.025 \\ 
\hline
\end{tabular}
\end{center}
\label{tabB}
\end{table}

\begin{table}
\centering
\begin{center}
\caption[]{The adopted folded $V$ light curve}
\begin{tabular}{ccc} \hline
Phase &    $V$ magnitude  &  magnitude error \\
\hline
0.003  &  14.861    &    0.020 \\ 
0.032  &  14.886    &    0.035 \\ 
0.094  &  14.859    &    0.036 \\ 
0.142  &  14.799    &    0.030 \\ 
0.203  &  14.674    &    0.024 \\ 
0.248  &  14.688    &    0.027 \\ 
0.301  &  14.676    &    0.047 \\ 
0.346  &  14.714    &    0.026 \\ 
0.409  &  14.840    &    0.037 \\ 
0.449  &  14.877    &    0.020 \\ 
0.505  &  14.916    &    0.022 \\ 
0.550  &  14.848    &    0.022 \\ 
0.602  &  14.777    &    0.030 \\ 
0.663  &  14.731    &    0.028 \\ 
0.703  &  14.720    &    0.022 \\ 
0.755  &  14.653    &    0.033 \\ 
0.798  &  14.681    &    0.030 \\ 
0.846  &  14.759    &    0.026 \\ 
0.896  &  14.862    &    0.024 \\ 
0.946  &  14.872    &    0.020 \\ 
\hline
\end{tabular}
\end{center}
\label{tabV}
\end{table}

\section{Ellipsoidal variations}

\subsection{Introduction and outline of model}

The folded light curves shown in Figure \ref{foldBV} 
have the well-known signature of
ellipsoidal modulations with maxima at phases 0.25 and 0.75 (the
quadrature phases) and minima at phases 0.0 and 0.50 (the conjunction phases).
The origin of ellipsoidal variations is easy to understand.  The secondary
star fills its critical Roche lobe (there is ongoing mass transfer) and hence
is greatly distorted.  As it moves around in its orbit its projected area on
the sky (and hence the total observed flux) changes.  If the secondary
star fills its Roche lobe and is in synchronous rotation, then the
amplitude of its ellipsoidal light curve is only a function of the orbital
inclination ($i=90^{\circ}$ for a system seen edge-on) and the binary mass
ratio.  Thus models of the ellipsoidal variations offer a way to 
measure the inclination and mass ratio.
However, the {\em observed} light curve may not be totally due to
ellipsoidal modulations from the secondary star.  
For example, the addition of a substantial amount
of extra light from the accretion disc
will reduce the overall observed
light curve amplitude.  X-ray heating of the secondary star can sometimes
radically alter
the shape of the observed light curve (Avni \& Bahcall 1975; Avni 1978).  
One must therefore account for
extra sources of light when modelling X-ray binary (optical) light curves.
The folded light curves shown in  Figure \ref{foldBV} 
show the secondary star in Cyg X-2
with the 
least contamination from the accretion disc (and possibly the least effects
of the irradiation).  
It has already been shown that using the lower envelope of the folded
light curve 
for inclination estimates is a good approximation, when compared to 
ellipsoidal variations with no disc or irradiation contamination
(e.g.\ Pavlenko et al.\ 1996).     We will discuss below models of the
folded light curves shown in Figure \ref{foldBV}
and the inclination (and hence mass)
constraints we derive from them.

We modeled the light curves using the
modified version of the
Avni (1978) 
code described in 
Orosz \& Bailyn (1997).  This code uses full Roche geometry
to describe the shape of the secondary star.  In addition, the code
accounts for light from
a  circular accretion disc and for extra light from the secondary star
due to X-ray heating. 
The parameters for the model are the parameters which determine
the geometry of the system:  the mass ratio 
$Q=M_x/M_c$\footnote{Following Avni, we use in this section the upper
case $Q$ to denote the mass ratio defined by the mass of the compact
star divided by the mass of the secondary star.  CCK98 and
others use the lower case $q$ to denote the inverse.}, the orbital
inclination $i$, the Roche lobe filling factor $f$, and the
rotational velocity of the secondary star;  the parameters which determine
the light from the secondary star:  its polar temperature $T_{\rm pole}$,  
the linearized limb darkening coefficients $u(\lambda)$, and the gravity
darkening exponent $\beta$;
the parameters which determine the contribution of
light from the accretion disc: the disc radius $r_d$, flaring angle
of the rim $\beta_{\rm rim}$, the temperature at the outer edge
$T_{d}$, and the exponent on the power-law radial
distribution of temperature $\xi$, where $T(r)=T_d(r/r_d)^{\xi}$;
and parameters related to the X-ray heating:  the X-ray luminosity of the
compact object $L_x$, the orbital separation (determined from the
optical mass function, the mass ratio, and the inclination), and
the X-ray albedo $W$.

For simplicity, we set many of the model parameters at reasonable values.
For example, we
assume the secondary star is in synchronous rotation and completely
fills its Roche lobe since there is ongoing mass transfer. 
We initially fix the mass ratio at $Q=2.94$, the value found by
CCK98.
Thus the only remaining geometrical parameter is the inclination
$i$.
Based on the secondary star's spectral type (A9III, CCK98)
we fix its polar temperature 
at 7000K.   We use limb darkening coefficients
taken from
Wade \& Rucinski (1985).  
The secondary star has a radiative
envelope, so the gravity-darkening exponent was set to 0.25
(von Zeipel 1924).

\subsection{Upper and lower bounds on the inclination}

We begin the light curve fitting by considering a simple model first.
We fit the $B$ and $V$ light curves separately to a model with no disc
light and no X-ray heating.  The only free parameter is the
inclination $i$.  In this case, the best derived value of $i$ is a
lower limit since the amplitude of the light curve gets smaller
as the inclination is decreased or if extra light from the accretion disc
is added.  We
find $i\ge 42^{\circ}$ for $B$ and $i\ge 49^{\circ}$ for $V$.  Thus we
adopt $i\ge 49^{\circ}$ as a lower limit.  The upper limit on $i$
based on the lack of X-ray eclipses is $i\le 73^{\circ}$ (for a mass
ratio of $Q=2.94$, see CCK98).  We therefore conclude
$49^{\circ} \le i \le 73^{\circ}$.

\subsection{Addition of disc light}

We now consider models where light from the disc
is accounted for.   The disc parameters discussed above specify the
disc radius, thickness, the temperature of the rim, and temperature profile.
At each phase,
the disc is divided up into a grid of 6300 surface elements (90 divisions
in azimuth times 70 divisions in the radial direction).
The local intensities are computed assuming backbody spectra
(with corrections for limb darkening)  at 128 different
wavelengths.
The observed flux at each wavelength
from the disc is obtained by summing over all of the visible surface elements,
creating a model disc ``spectrum.''  The model disc
spectrum is added to the model star spectrum
(constructed in a similar way
with local blackbody spectra and limb darkening
corrections), creating the ``composite''
spectrum.  This composite spectrum is integrated with the standard normalized
$UBVRI$ filter response functions to obtain the final model $UBVRI$
fluxes.    This method of computing the final fluxes is more precise than
simply computing the fluxes at the ``effective'' wavelengths of the filters
because the effective wavelengths of the filters depend on the
exact shape of the input spectrum.  This technique also
can easily handle cases where the disc spectrum has a
rather different shape than the stellar spectrum.

In
cases where one has observations in several filters, the value of
the radial exponent,
$\xi$ can be quite well constrained (Orosz \& Bailyn 1997).  However,
in the case of Cyg X-2 where we only have $B$ and $V$ light curves, we
found that the value of $\xi$ was not well constrained
(additional observations in $R$ and
$I$ would help constrain $\xi$, see below).  We therefore
initially
fixed $\xi$ at $\xi=-0.75$, which is the value appropriate for a 
steady-state disc (Pringle 1981).  

\subsection{Addition of X-ray heating}

We argue that even though Cyg X-2 is a strong X-ray source,
X-ray heating is relatively unimportant with
regards to the Cyg X-2 optical light curves.  It is well known that strong
X-ray heating of the secondary star in an X-ray binary can lead to
distortions in the secondary star's observed radial velocity curve
(e.g.\ Bahcall, Joss, \& Avni 1974) and to changes in the observed
spectral type of the secondary as a function of orbital phase
(e.g.\ Crampton \& Hutchings 1974).  However, in the case of Cyg X-2, 
no distortions in the radial velocity curve or changes
in the spectral type were observed by
CCK98.
X-ray heating of the secondary star
can also alter the observed optical light curve by adding light near the
photometric phase 0.5 (Avni 1978).  There is no evidence for any
significant excess light near $\phi=0.5$ in the folded light curves displayed
in Figure \ref{foldBV}.  
There are two main reasons why X-ray heating seems to be
unimportant in Cyg X-2.  First of all, Cyg X-2 has a much larger orbital
separation than other X-ray binaries such as Her X-1, so the X-ray flux
(i.e.\ the number of X-ray photons per unit surface area)
at the surface of the Cyg X-2 secondary star will be smaller.  Secondly,
many X-ray binaries are thought to have relatively thick accretion discs
(e.g.\ Motch et al.\ 1987;
de Jong, Augusteijn \& van Paradijs 1996).  
The thick discs may shield the secondary star
from much (if not most) of the X-rays from the central source. 
The accretion disc may also be warped (e.g.\ Wijers \& Pringle 1998),
in which case the secondary star may be shielded from the X-rays
from the central source.

Because X-ray heating is a small effect here, we will use a simplified
computational procedure.  
We assume that all of the X-rays come from a point centred
on the neutron star. 
The flux of X-rays on each point on
the secondary star that can see the central X-ray source is 
\begin{equation}
F_{\rm irr}=\Gamma {L_x\over d^2}, 
\label{firr}
\end{equation}
where $d$ is the distance between the point in question and the
centre of the neutron star, $L_x$ is the X-ray luminosity of
the central source, and where 
$\Gamma$ is cosine of the angle between the surface normal and the
direction to the central source. The rim of the accretion disc
can shield the points on the secondary star that are near the orbital
plane, preventing them from seeing the X-rays from the central source.
$F_{\rm irr}=0$ in these cases.  
The flux of X-ray photons on 
a particular surface element on the secondary star (specified
by the coordinates ($x,y,z$)) causes the local
temperature to rise according to
\begin{equation}
T^4_{\rm X-ray}(x,y,z)=T^4_{\rm pole}
       \left[ {g(x,y,z)\over g_{\rm pole}(x,y,z)}\right]^{4\beta}
       +{WF_{\rm irr}\over \sigma}.
\label{xrayeq}
\end{equation}
$W$ is the X-ray albedo, $g$ is the gravity,  and  $\sigma$
is the Stefan-Boltzmann constant (Zhang et al.\ 1986;
Orosz \& Bailyn 1997).
We see that the amount the local temperature is raised depends on the
value of the product $WF_{\rm irr}$.  Thus it is possible to derive different
values of $L_x$ by using different values of $W$---the final model light curves
are identical as long as the product $WF_{\rm irr}$ remains constant.
It is also possible to obtain larger values of
$L_x$ by using a thicker accretion disc, since thicker accretion discs
would shield larger parts of the secondary star.  However, in this case,
the final model light curves do have subtle differences. 
We adopt an X-ray albedo of $W=0.50$ definiteness
and initially  fix the
X-ray luminosity at $\log L_x=38.3$, which is roughly the Eddington
luminosity for a $1.5\,M_{\odot}$ neutron star. 

\begin{figure}
\centering
\centerline{\epsfxsize=9.0cm 
\epsfbox{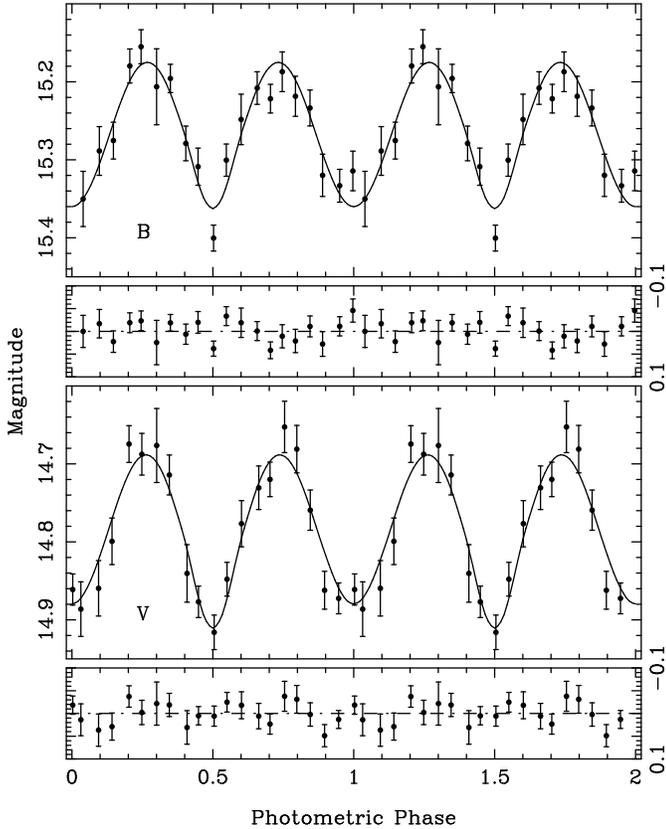}}  
\caption{From top to bottom:  The folded $B$ light curve and the
best-fitting model
(Table 3); the $B$ residuals in the sense of the data minus
the model fit; the folded $V$ light curve and the best-fitting model; and
the $V$ residuals.  Each point has been plotted twice.
}
\label{foldfig}
\end{figure}

\subsection{Light curve fitting and computation of confidence regions}

We fit the $B$ and $V$ light curves simultaneously.  
Models using a wide range of the input
parameters were computed and parameters sets which gave relatively low
values of $\chi^2$ were noted. 
Two different optimization routines were used: one based on the
``grid search'' algorithm  and one based on the
Levenberg-Marquardt algorithm (both adopted from Bevington 1969).
The optimization routines were then
started using each of these parameter sets as an initial guess. Our
best-fitting model (arrived at after several weeks of computation) is
shown in Figure \ref{foldfig}.  Table 3 gives the values of
the free parameters (which are the inclination $i$, the radius
of the disc as a fraction of the Roche lobe radius $r_d$,
the opening angle of the disc $\beta_{\rm rim}$, and the temperature
of the outer edge of the disc $T_d$)
The model fits reasonably well, with
$\chi^2=40.93$ for 36 degrees of freedom.  The standard deviations of the
residuals are 0.024 mag for $B$ and 0.026 mag for $V$.  We are reasonably
certain the global $\chi^2$ minimum was found, based on the large amount
of parameter space searched.

\begin{table}
\centering
\begin{center}
\caption[]{The model parameters for the light curves shown in Figure
\protect\ref{foldfig} (steady-state disc case)}
\begin{tabular}{ccc} \hline
Parameter &    Value  &   Comment \\
\hline
$Q$      &  2.94           &    fixed  \\
$i$      &  $62.5^{\circ}\pm 4^{\circ}$ &    free   \\
$r_d$    &  $0.90\pm 0.15$        &    free   \\
$\beta_{\rm rim}$  & $15.04^{\circ}\pm 0.07^{\circ}$  & free \\
$T_d$ (K)   &  $3950 \pm 880$         &    free   \\
$\xi$    &  $-0.75$        & fixed     \\
$W$      &   0.5           & fixed     \\
$\log L_x$ (cgs)  &  38.3      & fixed      \\ \hline
\hline
\end{tabular}
\end{center}
\label{tab1}
\end{table}

\begin{figure}
\vspace{6cm}
\includegraphics{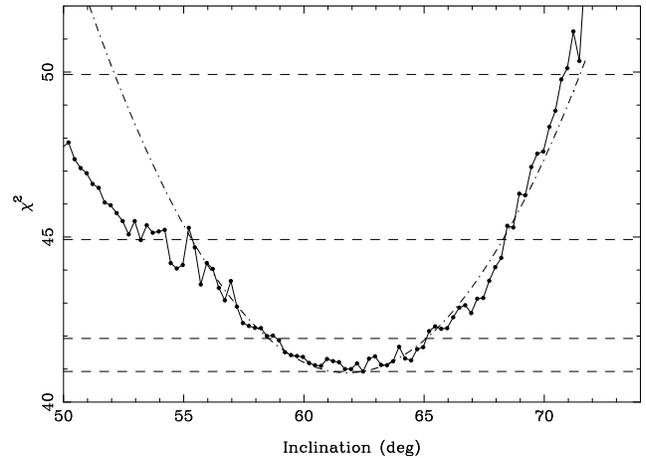}
\caption{$\chi^2$ vs.\ $i$ for a mass ratio of $Q=2.94$
(solid line and filled circles).
The horizontal dashed lines correspond to $\chi^2=\chi^2_{\rm min}$
$\chi^2=\chi^2_{\rm min}+1$, and $\chi^2=\chi^2_{\rm min}+4$,
and $\chi^2=\chi^2_{\rm min}+9$.  The dash-dotted line is a parabola
fitted to the $\chi^2$ values between $55^{\circ}$ and $70^{\circ}$.
}
\label{crosscut}
\end{figure}

The parameter errors given in Table 3 were computed with the
Levenberg-Marquardt optimizer program.  
The sizes of the errors depend on the sizes of the parameter
increments one uses to compute the numerical derivatives.
As a check on these errors, we
also estimated $1\sigma$ and $2\sigma$ confidence regions in the
following way.  A grid in the inclination-mass ratio plane was defined
with inclinations between $49^{\circ}$ and $74^{\circ}$ in steps of
$0.25^{\circ}$ and mass ratios of 2.66, 2.94 and 3.22.
Then the light curves were fit with the inclination and mass ratio
fixed at the values corresponding to each grid point.  In each case,
the other parameters were allowed to vary until $\chi^2$ was
minimized.  This iteration was started at the point where
$i=62.46^{\circ}$ and $Q=2.94$.  Then the fit was optimized at a
neighbour point using the parameters at the lowest
neighbour point as the
initial guess for the optimization routine.  The iteration was
continued until the entire grid was filled up.  We found that the
value of $\chi^2$ did not depend strongly on the mass ratio for a
given inclination.  We show in Figure \ref{crosscut} $\chi^2$ vs.\ $i$
for $Q=2.94$ (the solid line and filled points).
We also show a parabolic fit to the $\chi^2$ values between $55^{\circ}$
and $70^{\circ}$ (dash-dotted line).  This fit shows that the
$\chi^2$ vs.\ $i$ curve
is roughly parabolic near the minimum.
We see that $\chi^2=\chi^2_{\rm
min}+1$ at $i\approx 59^{\circ}$ and at $i\approx 65^{\circ}$.  Thus
$59^{\circ}\le i \le 65^{\circ}$ is an approximate $1\sigma$
confidence region.  A rough $2\sigma$ confidence region is
$56^{\circ}\le i \le 68^{\circ}$ where $\chi^2=\chi^2_{\rm min}+4$ at
the endpoints.  
The value of $\chi^2$ increase sharply as
the inclination grows beyond $\approx 68^{\circ}$ since the model predicts
deep eclipses that are not observed.  
The rough $1\sigma$ errors of $\pm 3^{\circ}$ are not too different 
from the $1\sigma$ errors of $\pm 4^{\circ}$ computed from the 
Levenberg-Marquardt program.  We will adopt $1\sigma$ errors of
$\pm 4^{\circ}$ for the sake of discussion below.

\begin{figure}
\centering
\centerline{\epsfxsize=6.0cm 
\epsfbox{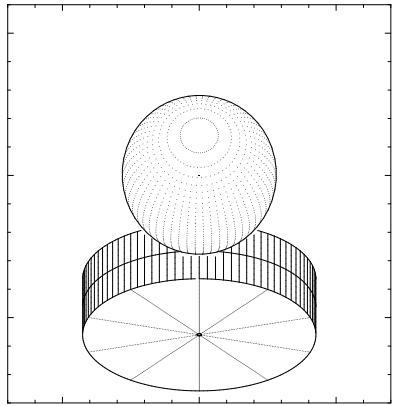}}  
\caption{A schematic of the binary system as it appears on the plane
of the sky at phase 0.0 for the model parameters given in Table 
3.}
\label{schem}
\end{figure}

\begin{figure}
\centering
\vspace{6cm}
\includegraphics{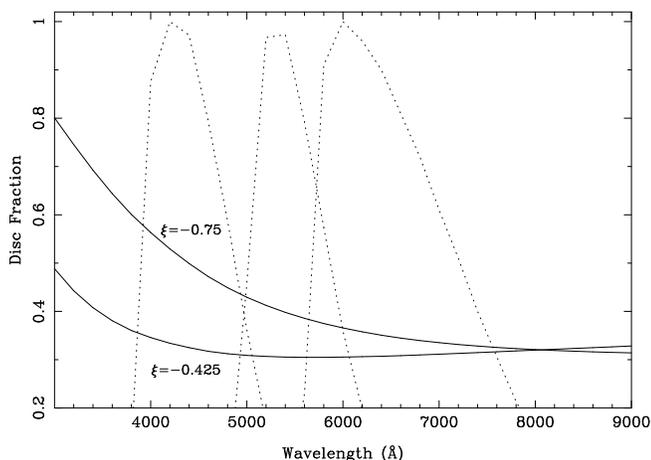}
\caption{The disc fraction as a function of wavelength for the
$\xi=-0.75$ model and for the $\xi=-0.425$ model (solid lines).  
The dotted lines show the filter response functions for the standard
$B$, $V$, and $R$ filters.
 }
\label{diskfrac}
\end{figure}

\subsection{The system geometry: A grazing eclipse of the disc?}

We show in Figure \ref{schem} a schematic diagram of the binary system
(using the model parameters given in Table 3)
as it appears on the plane of the sky at phase 0.0.  In this geometry
we expect a grazing eclipse of the disc.
The
grazing eclipse of the disc
results in a $V$ light curve that is $\approx 0.02$ mag
deeper at the photometric phases 0.0 and 0.5
than the uneclipsed light curve.  In principle, one
would expect to observe characteristic changes in the disc emission line
profile (e.g.\ H$\alpha$) as a function of phase for eclipsing systems (Young
\& Schneider 1980).  However, it is not clear if the predicted grazing
eclipse in Cyg X-2 is deep enough to lead to easily observable
changes in the H$\alpha$ line profile.  Future spectroscopic observations
at the correct orbital phases may help to further constrain the inclination
if one could show that partial eclipses do or do not occur.

\subsection{Changes in the disc temperature profile and
secondary star temperature}

We did some numerical experiments to explore how the model fits depend
on the parameter $\xi$.  The reason for these experiments is that
the disc in Cyg X-2 is probably strongly
irradiated by the central X-ray source, and it is likely that
the temperature profile of the disc is changed.  
Vrtilek et al.\ (1990) showed
that the temperature profile of an irradiated accretion disc
is like $T(r)\propto r^{-3/7}$ rather than the familiar
$T(r)\propto r^{-3/4}$ for a steady-state non-irradiated
disc.
[Recently, Dubus et al.\ (1998) argued that a 
non-warped disc irradiated by a
point X-ray source
powered by accretion 
is unchanged by the irradiation.
The observation that the accretion discs in
most LMXBs are effected by irradiation leads Dubus et al.\ (1998) to
conclude that the discs are either warped or that the central X-ray
source is not point-like.]
We computed models where the parameter $\xi$ was fixed
at several values between $-0.425$ and $-0.750$ and where the
other parameters were adjusted to minimize $\chi^2$ (the mass ratio
was fixed at $Q=2.94$).    
As $\xi$ moves from $-0.75$  to 
$-0.425$, the value of $\chi^2$ of the fit increases and the best-fitting
value of the inclination $i$ decreases.  When $\xi=-3/7\approx -0.425$, 
$i= 54.6^{\circ}$  and $\chi^2=48.4$ for 36 degrees of freedom.
We did not estimate confidence regions for the inclination
for the case when
$\xi=-3/7$.

We also computed models where the polar temperature of the secondary
star was slightly altered from its nominal value of 7000~K.  When
$T_{\rm pole}=6700$~K
(corresponding roughly to a spectral type of
F2), the best-fitting inclination was 
$i=63.6^{\circ}$ with $\chi^2=40.90$ for 36 degrees of freedom, slightly
lower than the value of 40.93 found above for the model with
$T_{\rm pole}=7000$~K.  
When
$T_{\rm pole}=7400$~K
(corresponding roughly to a spectral type of
A7), the best-fitting inclination was 
$i=61.5^{\circ}$ with $\chi^2=41.35$ for 36 degrees of freedom.  
Thus our results do not depend strongly on the adopted value of the
secondary star's polar temperature.

\subsection{Comparison of observed and model disc fraction}

We have an independent check on the light
curve  models.  Since the spectrum of
the star and the spectrum of the disc are computed separately, we can
easily predict what fraction of the flux at a given wavelength is due to the
disc
(we refer to this number as the ``disc fraction'' and denote it by $k$). 
Observationally,
the disc fraction in a given
bandpass can be measured using the optical spectrum of the source and the
spectra of suitable template stars (Marsh, Robinson, \& Wood 1994).
We obtained two spectra of Cyg X-2 1997 on November 1 and 3 with the
2.7m telescope at the McDonald Observatory (Fort Davis, Texas)
using the Large Cass Spectrograph, a 600 groove/mm grating 
(blazed at 4200~\AA), and the TI1 $800\times 800$ CCD.  The
spectral resolution is $\approx 3.5$~\AA\ (FWHM) with wavelength
coverage 3525-4940~\AA.  The signal-to-noise ratios are $\approx 50$
per pixel near the He II emission line at 4686~\AA.
We also observed the A9III star
HR 2489 on both nights.  CCK98 found that the spectrum
of this star best matched the absorption line spectrum of Cyg X-2.  We 
normalized each spectrum to its continuum fit and applied the technique
of Marsh, Robinson, \& Wood (1994) to decompose the spectra of Cyg X-2
into the disc and stellar components.  We found a disc fraction of 
$k_B=0.36\pm 0.05$ from the November 1 spectrum 
(orbital phase $=0.65$) and $k_B=0.37\pm 0.05$ for the
November 3 spectrum
(orbital phase $=0.85$).  We note that these estimates of $k_B$ 
represent an average value over most of the $B$ bandpass.
For comparison, J. Casares (private communication) 
finds from the much higher quality spectra published in CCK98
that the disc fraction in the $B$ and $R$ bands was variable, with
$0\la k_B\la 0.30$ and $0\la k_R \la 0.40$ for the $B$ and $R$ bands,
respectively.  
For this discussion we
adopt $k_B=0.30$ and $k_R=0.30$.

We show in  Figure \ref{diskfrac} the model disc
fraction
as a function of wavelength for two models:  the best-fitting solution
with $\xi=-0.75$ and the best-fitting solution with $\xi=-0.425$.
We also show the standard $B$, $V$, and $R$ filter response functions.
The disc fractions for the $\xi=-0.75$ model are $k_B\approx 0.55$ and
$k_R\approx 0.35$.  Both of these values are somewhat larger than 
the observed values.  The disc fractions in $B$ and $R$
for the $\xi=-0.425$
model are slightly larger than $\approx 0.3$, much closer to the 
upper ranges of the observed
values.  However, the fit for the $\xi=-0.425$ model is worse than the
fit for the $\xi=-0.75$ model ($\chi^2=48.44$ compared to $\chi^2=40.92$
for 36 degrees of freedom).  

\subsection{The puzzling lack of X-ray heating}

We have argued above that X-ray heating of the secondary star in Cyg X-2
is not a large effect since there is no observed change in the spectral
type as a function of orbital phase and because there does not appear
to be a large amount of excess light at the photometric phase 0.5.  
The effect of the X-ray heating in the model given in Table 3 is
small---there is about $\approx 0.01$ to 0.02 magnitudes of excess light
added near phase 0.5 compared to the light curve without
heating.
This is in spite of the fact that the neutron star is accreting at
nearly the Eddington rate (see Smale 1998).  
In our model the
rim of the disc shields most of the secondary star from the
central X-ray source [although it is also likely that the disc is warped
(Dubus et al.\ 1998), our model currently only uses an axisymmetric disc].

It turns out that solving the ``problem'' of no X-ray heating of the
secondary star leads to another puzzle.  Namely, it 
is it clear
observationally that the disc in Cyg X-2
is slightly fainter in the optical than the secondary star
(i.e.\ the observed disc fractions in $B$ and $R$ are $\approx 0.3$).
However, 
Jan van Paradijs pointed out to us that 
relative faintness of the disc is somewhat surprising since
one would expect a substantial 
amount of light from the reprocessing of the X-rays
absorbed by the disc.  The He II $\lambda4868$~\AA\ line is in emission,
so presumably there is at least some X-ray reprocessing.
Based on the relations given in van Paradijs \& McClintock
(1994), the absolute $V$ magnitude of the accretion disc
should be $M_V=-2.02\pm 0.56$.  We compute $M_V=0.38\pm 0.35$ 
for the disc, based on 
our model parameters (the secondary star by itself has $M_V=-0.54
\pm 0.24$, see Section 4.2).  Thus the disc in Cyg X-2 is about
a factor of 9 fainter in $V$ than expected based on the simple scaling
laws given in van Paradijs \& McClintock (1994).   
It is possible that 
much of the reprocessed flux from the disc is
emitted at shorter wavelengths than $B$, which might account for some
of the mismatch between the observed and expected disc brightness.

Presently,  the code does not self-consistently account for 
optical flux from the disc due to reprocessing of absorbed X-rays.
The brightness of the disc is set based on the temperature at the
outer edge, the radial temperature profile, and the disc radius,
and is independent on the adopted value of $L_x$.  
The code is ``flexible'' in the sense that 
the power-law exponent on the temperature radial profile can be adjusted
to approximate the changes in the disc caused by
irradiation.
As we showed above,
the model disc fractions are not too different than what is observed.
It therefore appears that this lack of a self-consistent computation 
of the disc flux is not a major problem.

\subsection{Summary of model fitting and discussion}

The fits to the $B$ and $V$ light curves are better
if we assume that the disc is in a steady state where
$T(r)\propto r^{-0.75}$.  However, in this case the model disc is bluer than
what is observed since our predicted disc fraction in $B$ ($k_B\approx
0.55$) is
larger than what is observed ($k_B\approx 0.30$).  If
we assume that the disc is strongly irradiated so that  
$T(r)\propto r^{-0.425}$, then the model disc is redder and the 
predicted disc fraction in $B$ ($k_B\approx 
0.32$) is closer to what is observed.
However, in this case, the $\chi^2$ of the fit is much worse 
($\chi^2=48.4$ compared to $\chi^2=40.9$ with 36
degrees of freedom for the steady-state
case).  Fortunately, the best-fitting values of the inclination are not
that different in the two cases: $i\approx 62.5^{\circ}$ for the steady-state
disc case and $i\approx 54.6^{\circ}$ for the irradiated case, which is in the
$2\sigma$ range of the steady-state disc case.
For the sake of the discussion in Section 4
we adopt the $1\sigma$ inclination from the
steady-state case ($i=62.5^{\circ}\pm 4^{\circ}$,
where we assume the errors are Gaussian) because of
the lower $\chi^2$.  

It is important to recall here that observed optical light curves
consist mainly of two components:  the light from the distorted secondary
star and the light from the accretion disc.  We have argued that there
is very little extra light due to X-ray heating of the secondary star. 
We assume that the light from the secondary star is modulated in phase
while the light from the disc is not
(with the possible exception of a grazing eclipse).  
Thus to model the observed light
curves we should compute the relative amounts of disc light and secondary
star light at each observed wavelength region
for every observed phase.  

In our current model we compute the disc light at every observed wavelength
region by specifying four parameters: the disc radius in terms of the
neutron star Roche lobe radius, the opening angle of the disc rim,
the temperature profile of the disc, and the temperature of the disc rim.
The flux at each grid point across the disc is computed from the local
temperature assuming a blackbody spectrum. 
The code does not account for flux from the
disc due to reprocessing of absorbed X-rays
from the central source.
Typically, the spectrum of the disc (in the optical) will have a rather
different shape than the spectrum of the secondary star.  Hence
one should have observed light curves in as many wavelength
bands as possible in order to better determine the shapes of the disc 
and stellar spectra.  
Eclipsing systems with well-defined and smooth light curves
like GRO J1655-40
(Orosz \& Bailyn 1997; van der Hooft et al.\ 1998) offer additional
constraints on the disc radius, thickness, and temperature.  

On the other hand, there is no particular reason to adopt our parameterization
of the disc since the important quantity is the
relative amount of disc and star light at a particular wavelength. 
In fact, by using suitably high quality spectra
one could simply measure the disc fraction $k_{\lambda}$ at
several different wavelengths covering the bandpasses of the observed light
curves.  In this case the model disc spectrum would be constructed 
from the model star spectrum since the quantity $k_{\lambda}=
f_{\rm disc}/(f_{\rm disc}+f_{\rm star})$ is known an each wavelength point
(here $f_{\rm disc}$ and $f_{\rm star}$ refer to the model fluxes from
the disc and star, respectively, at a given orbital phase).
Then, as before, the model disc spectrum is added to the model star
spectrum and the resulting composite spectrum is integrated with the
filter response curves to produce model fluxes in each bandpass.  
Thus one could eliminate the model parameters $T_d$
and $\xi$.  
We note that  one still needs a model disc to account
for the effects of X-ray shadowing by the disc rim and possibly the 
slight loss
of flux due to the eclipse, so the model parameters $\beta_{\rm rim}$
and $r_d$ are still needed.

Thus, future modelling of the Cyg X-2 light curves can be improved by
observing the light curves in more colours (i.e.\ at least in
$B$, $V$, $R$, and $I$ and possibly also in the infrared), and by 
obtaining quasi-simultaneous spectroscopic observations of
Cyg X-2 and template stars over a wide wavelength range.  In practice,
one needs  observations over many orbital cycles in order to average out
the variations in the observed disc fraction and to define the lower
light curve envelopes.   

\section{Discussion}

\subsection{Mass of the neutron star}

The masses of the neutron star and secondary star can be
computed from the optical mass function,
the mass ratio, and the inclination (CCK98):
\begin{equation}
M_x={f(M)(1+q)^2\over\sin^3 i}=(1.24\pm 0.09\,M_{\odot})(\sin i)^{-3}.
\end{equation}
Using our $1\sigma$ limits on the inclination 
($i=62.5^{\circ}\pm 4^{\circ}$)
we find $M_x=1.78\pm 0.23\,M_{\odot}$
and $M_c=0.60\pm 0.13\,M_{\odot}$. The extreme range of allowed
inclinations ($49^{\circ}\le i\le 73^{\circ}$) implies an
extreme mass range
allowed for the neutron star of $1.42\pm 0.10\,M_{\odot}
\le M_x\le 2.88\pm 0.21\,M_{\odot}$.

\begin{figure}
\vspace{6cm}
\includegraphics{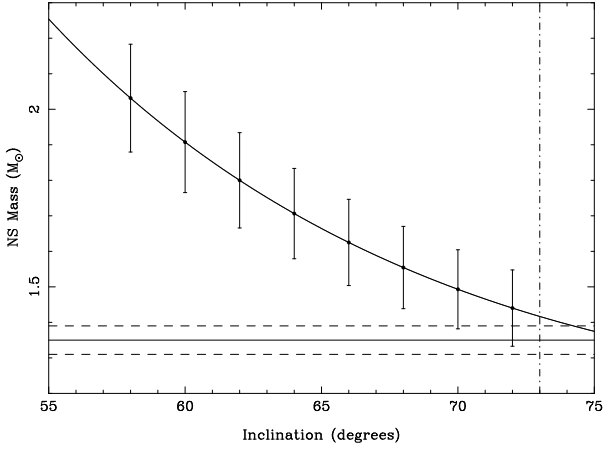}
\caption{The mass of the neutron star as a function of the inclination
is shown.  We assume an optical mass function of $0.69\pm 0.03$
and a mass ratio of $M_c/M_x=0.34\pm 0.04$ (CCK98).  The vertical
line indicates the upper limit on the inclination imposed by the lack
of observed X-ray eclipses.  The horizontal lines indicate the 
canonical neutron star mass of $1.35\pm
0.04\,M_{\odot}$ found from the binary
radio pulsars.
}
\label{massplot}
\end{figure}

Since the values of the optical mass function and the mass ratio were
fairly well determined by CCK98, the largest uncertainty on $M_x$ is
the value of the inclination $i$ one chooses.  Thus it is instructive
to see how the mass of the neutron star varies as a function of the
inclination.  We plot in Figure \ref{massplot} the mass of the neutron
star as a function of the inclination.  We also indicate the $1\sigma$
errors at several different inclinations.  The mass of the neutron
star in Cyg X-2 is consistent at the $1\sigma$ level with the
canonical neutron star mass of $1.35_{\odot}$ (Thorsett \& Chakrabarty
1998) for inclinations greater than $70^{\circ}$.  However, the fits
to the light curves get increasingly worse as the inclination grows
larger than $\approx 68^{\circ}$ (i.e.\ the sharp increase in $\chi^2$
displayed in Figure \ref{crosscut}), so it is unlikely that the
inclination of Cyg X-2 is much larger than $\approx
68^{\circ}$.  At the lower
value of the $1\sigma$ inclination range for our model with the
unirradiated disc ($i=58.5^{\circ}$), the neutron star mass is
$M_x=2.00\pm 0.15\,M_{\odot}$, which is more than $4\sigma$ larger than
the canonical mass of $1.35\,M_{\odot}$.

For most of the parameter space in $i$, $f(M)$, and $q$, the mass
of the neutron star in Cyg X-2 exceeds the average mass of the
neutron stars in binary radio pulsars.  Thus Cyg X-2 contains a rare
example of a ``massive'' neutron star.  Perhaps the best-known
example 
of a massive neutron
star is the high-mass
X-ray binary Vela X-1 (4U 0900-40).  Vela X-1 is an
eclipsing system, and the neutron star is an X-ray pulsar.
Dynamical mass measurements
by van Kerkwijk et al.\ (1995b)
gave a mass of $M_x=1.9^{+0.7}_{-0.5}$ (95\% confidence limits).  
A later  analysis of
IUE spectra by Stickland et al.\ (1997) gave a mass consistent
with $1.4\,M_{\odot}$
($1.34\,M_{\odot}
\la M_x\la 1.53\,M_{\odot}$).  However, a recent reanalysis of the UV
data (Barziv et al.\ 1998, in preparation) 
gives $M_x=1.9\,M_{\odot}$, consistent with the result of
van Kerkwijk et al.\ (1995b).
Barziv et al.\ (1998, in preparation) 
also find $M_x=1.9\,M_{\odot}$ from optical spectra.  The
range of derived masses by different groups is an indication of
the difficulty in analyzing the radial velocity curves of a high-mass
companion star such as the one in Vela X-1.  
Another example of a possible massive neutron star is the eclipsing
high-mass
X-ray binary
4U 1700-37.  The companion star (HD 153919) is an O6f star with a strong
wind.  Heap \& Corcoran (1992) find $M_x=1.8\pm 0.4\,M_{\odot}$.
However, 4U 1700-37 does not pulse and its X-ray spectrum is harder than
the spectra of typical X-ray pulsars.  This lack of ``neutron star
signatures'' has led Brown, Weingartner, \& Wijers (1996)
to speculate that 4U 1700-37 contains a ``low-mass'' black hole rather
than a neutron star. 

Recent computations of the neutron star and
black hole initial mass function by Timmes, Woosley,
\& Weaver (1996) indicate that Type II supernovae give rise
to a bimodal distribution of initial neutron star masses.
The average masses for the two peaks are $1.26\pm 0.06 \,M_{\odot}$
and $1.73\pm 0.08\,M_{\odot}$, respectively. Type
Ib supernovae tend to produce neutron stars in the lower mass
range.
The masses derived by Timmes et al.\ (1996) do not include mass
that may fall back onto the neutron star shortly after the
supernova explosion.  The mean mass for the lower-mass distribution
of $1.26\pm 0.06\,M_{\odot}$ agrees well with the mean mass of
the binary radio pulsars of $1.35\pm 0.04\,M_{\odot}$ determined
by Thorsett \& Chakrabarty (1998).  The mass of the neutron star in Cyg X-2
seems to be significantly larger than both of these masses.
However, the neutron star mass of $M_x=1.78\pm 0.23
\,M_{\odot}$ given above agrees well with the mean mass of
$1.73\pm 0.08\,M_{\odot}$ derived by Timmes et al.\ (1996)
for the higher-mass peak of their bimodal distribution. Thus the current
mass of the neutron star in Cyg X-2 might simply be the mass
at its formation
(within the framework
of the models of Timmes et al.\ (1996)). 
%
Alternatively, 
Zhang, Strohmayer, \& Swank
(1997) point out that 
one would expect massive neutron
stars to exist in systems where a neutron star formed at 
$\approx 1.4\,M_{\odot}$
has been accreting at the Eddington rate for extended periods
of time.
If the kilohertz QPOs observed in Cyg X-2 and other
neutron star LMXBs can be interpreted as the frequency of the
last stable orbit of the inner accretion disc, then the neutron
star masses in some X-ray binaries
could be as large as $\approx 2\,M_{\odot}$.  Assuming the neutron
stars were formed at $\approx 1.4\,M_{\odot}$, the $\approx 0.6\,M_{\odot}$
of extra matter is not an unreasonable amount to accrete 
in $\approx 10^8$ years (Zhang et al.\ 1997).  
One would basically have to know how long Cyg X-2 has been 
accreting at near Eddington rates in order to determine whether the
neutron star formed at ``low mass'' ($\approx 1.3\,M_{\odot}$) or
``high mass'' ($\approx 1.7\,M_{\odot}$).  It remains to be seen if
a reliable age estimate can be derived from a binary evolution model
of this system.

According to King et al.\ (1997), a companion star mass of at least
$\approx 0.75\,M_{\odot}$ is needed to maintain steady accretion in a
neutron star low-mass X-ray binary like Cyg X-2.  If $M_c>
0.75\,M_{\odot}$, then $M_x> 1.88\,M_{\odot}$ at the 95\% confidence
level (CCK98), which would require an inclination lower than about
$60^{\circ}$, near the lower end of the $1\sigma$ inclination range.
We find
$M_c=0.60\pm 0.13\,M_{\odot}$ ($1\sigma$) using $i=62.5^{\circ}\pm
4^{\circ}$, which is $\approx 1\sigma$ smaller than minimum $M_c$ of
King et al.\ (1997).  The extreme range of allowed inclinations
($49^{\circ}\le i\le 73^{\circ}$) implies an extreme mass range
allowed for the secondary star of $0.48\pm 0.09\,M_{\odot} \le M_c\le
0.98\pm 0.18\,M_{\odot}$.  In principle, a measurement of the surface
gravity of the secondary star would provide an independent estimate of
$M_c$ since the density of the Roche-lobe filling secondary star in a
semi-detached binary depends only on the orbital period to a good
approximation (Pringle 1985).  However, one would need to measure
$\log g$ to better than $\approx 0.03$ dex to distinguish between
$M_c=0.60\,M_{\odot}$ and $M_c=0.75\,M_{\odot}$.


\subsection{The Distance to the Source}

We can compute the distance to the source using the results of our
model fitting.  Once the inclination $i$ is given, we can compute the total
mass of the system.    
The size of the semimajor axis $a$ is then computed from Kepler's third
law.  We then use
Eggleton's (1983) formula to compute
the effective radius of the secondary's Roche lobe in terms
of the orbital separation $a$:
\begin{equation}
{R_{Rl}\over a}={0.49q^{2/3}\over 0.6q^{2/3}+\ln(1+q^{1/3})}.
\end{equation}
The intrinsic luminosity of the secondary star then follows from
the Stefan-Boltzmann relation, where we assume $T_{\rm eff}=7000\pm 250$~K
and a bolometric correction of $0$.
To get the intrinsic luminosity of the entire system we must add light
from the accretion disc.  We do not
have spectroscopic observations in the $V$ band available so
we will interpolate between the measurements of
$k_B$ and $k_R$ and adopt $k_V=0.30\pm 0.05$.
Finally, the distance modulus can be computed
after we account for interstellar extinction.  The most complete
study of the interstellar extinction in the direction of Cyg X-2
is that of McClintock et al.\ (1984).  They derived a colour
excess of $E(B-V)=0.40\pm 0.07$ based on spectra from the
{\em International Ultraviolet Explorer} and on optical photometry
and spectra of 38 nearby field stars.  Assuming $A_V=3.1E(B-V)$,
the $1\sigma$ $A_V$ range of McClintock et al.\ (1984) is
$1.02\le A_V\le 1.46$.  Using this $A_V$ range,
we find a distance of
$d=7.2\pm 1.1$ kpc, where we have adopted $i=62.5^{\circ}\pm 
4^{\circ}$ and a
mean apparent $V$ magnitude of 14.8 for the ``quiescent'' state
(Figures \ref{foldBV}, \ref{foldfig}).  The absolute
$V$
magnitude of the system is
$M_V=-0.93\pm 0.25$, and the absolute $V$ magnitudes of the components
separately are 
$M_V=0.38\pm 0.35$ 
for the disc, and $M_V=-0.54
\pm 0.24$ for the secondary star, respectively.  Finally,
if we use the inclination derived using the irradiated disc
($i=54.6^{\circ}$), then we find a distance of $d=7.9$ kpc.

\begin{figure}
\centering
\centerline{\epsfxsize=9.0cm 
\epsfbox{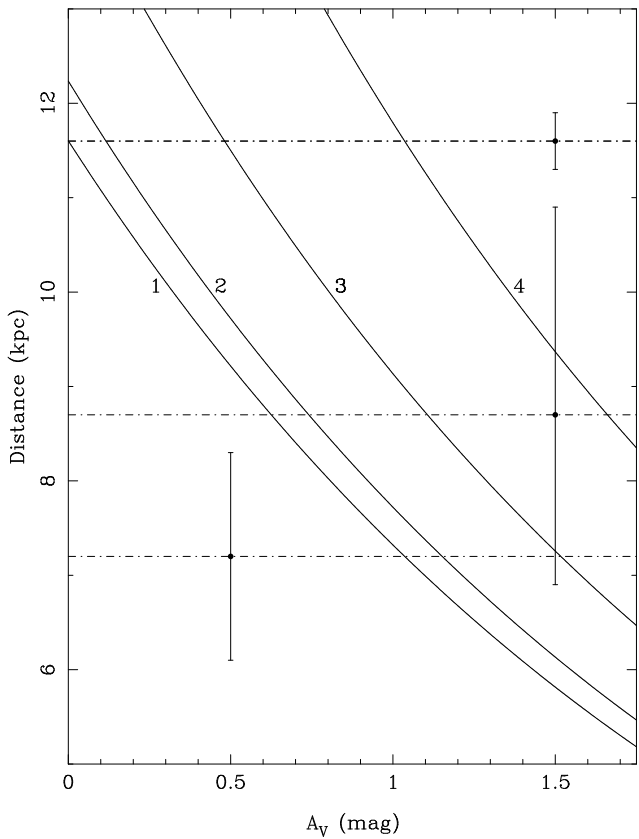}}  
\caption{The distance to Cyg X-2 as a function of the visual extinction
$A_V$.   The numbered curves show the
distance for four combinations of the inclination
and disc fraction:
(1)~$i=70^{\circ}$, $k_V=0.3$;
(2)~$i=63^{\circ}$, $k_V=0.3$;
(3)~$i=63^{\circ}$, $k_V=0.5$; and
(4)~$i=63^{\circ}$, $k_V=0.7$.  The three points with the error bars
indicate the distance derived by us
($d=7.2\pm 1.1$ kpc), by Cowley et al.\ (1979, $d=8.7^{+2.2}_{-1.8}$ kpc),
and by Smale (1998, $d=11.6\pm 0.3$ kpc).  The locations of the
points along the $A_V$-axis have no particular meaning.}
\label{distanceplot}
\end{figure}

Previous distance estimates include the 
distance derived by Cowley et al.\ (1979) of
$d=8.7\pm 2.0$~kpc, where we have propagated 
their estimated error on the bolometric magnitude of $\sim$0.5~mag (see Smale
1998). 
Our distance estimate is consistent with that of Cowley et al.\
(1979).
Recently, Smale (1998) derived a distance of $11.6\pm 0.3$~kpc from
observations of a type I
radius-expansion burst in Cyg X-2, where
he assumed a neutron star mass of 1.9~M$_{\odot}$ 
to derive the Eddington luminosity of the neutron
star.  Our derived distance is $\approx 4\sigma$ smaller than that of
Smale (1998).    

To facilitate a comparison between the various distance estimates, we
plot in Figure \ref{distanceplot} the distance to Cyg X-2 as a function
of the visual extinction $A_V$.  Since the distance to the source
depends weakly on the assumed inclination and somewhat strongly
on the assumed disc fraction, we show four $d$ vs.\ $A_V$ curves:
(1)~$i=70^{\circ}$, $k_V=0.3$;
(2)~$i=63^{\circ}$, $k_V=0.3$;
(3)~$i=63^{\circ}$, $k_V=0.5$; and
(4)~$i=63^{\circ}$, $k_V=0.7$.  The computed distance to the source
decreases sharply as $A_V$ grows larger.
If the
$1\sigma$ $A_V$ range of McClintock et al.\ (1984) is correct
($1.02\le A_V\le 1.46$), then the disc fraction must be greater
than $k_V\ga 0.7$ in order for the derived distance to be consistent
with Smale's (1998) measurement.  However,
a disc fraction of $k_V=0.7$ is much larger than what is observed:
our spectra and
the spectra of CCK98 give disc fractions of $\la 0.3$ in the $B$
and $R$ bands.  If the disc fraction of $k_V=0.3$ is correct,
then the visual extinction must be rather small ($A_V\la 0.2$ mag)
in order to get a distance of $\approx 11$ kpc.  
However, we note that the distance derived by Smale (1998) depends on
the assumed neutron star mass
(and other parameters, e.g.\ Lewin, van Paradijs, \&\ Taam 1993).  
If one assumes a
neutron star mass of $M_x=1.78\,M_{\odot}$,    
the distance would be $d=11.2\pm 0.3$ kpc (using standard
burst parameters).  This distance is marginally consistent with 
the estimate of Cowley et al.\ (1979), but still inconsistent with
our distance estimate.  One needs a rather small neutron star mass
($M_x\la 0.9\,M_{\odot}$) in order to get a distance from
the type I radius-expansion burst consistent with our
measurement.
Thus it is quite difficult to reconcile the differences between our
distance estimate and that of Smale (1998).

\section{Summary}

We have collected from the literature $U$, $B$, and $V$ light curves
of Cyg X-2.  The $B$ and $V$ light curves show significant periodicities
in their power spectra.  The most significant periodicities
in the $B$ and $V$ light curves correspond to half of the orbital period
of $P=9.8444$ days.  
The ``quiescent'' light curves derived from the lower envelopes of the
folded $B$ and $V$ light curves are ellipsoidal.  We fit ellipsoidal
models to the ``quiescent'' light curves;
from the best-fitting model we derive a $1\sigma$ inclination
range of $i=62.5\pm 4^{\circ}$, 
and a lower limit on the inclination of $i\ge 49^{\circ}$.  
The mass of the neutron star is $M_x=1.78\pm 0.23\,M_{\odot}$, where
we have used previous determinations of the
mass ratio
($q=M_c/M_x=0.34\pm 0.04$) and the optical mass function
($f(M)=0.69\pm 0.03\,M_{\sun}$), and the
$1\sigma$ inclination range of $i=62.5^{\circ}\pm 4^{\circ}$.
We find a distance of $d=7.2\pm 1.1$ kpc, which is significantly smaller
than a recent distance determination of $d=11.2\pm 0.3$ kpc derived
from an observation of a type I radius-expansion X-ray burst
(assuming $M_x=1.78\,M_{\odot}$),
but consistent with earlier estimates.

\section*{Acknowledgments}

This research has made use of the Simbad database, operated at CDS, 
Strasbourg, France.  We thank Tariq Shahbaz,
Jan van Paradijs, Phil Charles, and Lex Kaper 
for various useful
discussions and Jorge Casares for making his measurements of the disc
fraction available to us.

\bsp

\label{lastpage}

\end{document}